\documentclass[aps,prl,amsmath,amssymb,showkeys,twocolumn,superscriptaddress]{revtex4}
\usepackage{amsmath}%
\usepackage{amsfonts}%
\usepackage{amssymb}%
\usepackage{graphicx}%
\usepackage{color}
\definecolor{orange}{rgb}{1,0.5,0}
\definecolor{blue}{rgb}{0,0,1}
\usepackage{upgreek}

\usepackage{setspace}
\begin{document}
%\title[Physical Fractals and Apollonian Packing]{Physical Fractals and Apollonian Packing}
\title{Apollonian packings as physical fractals}

\author{Francesco~Varrato}
\affiliation{Ecole Polytechnique F\'ed\'erale de Lausanne (EPFL), Institute of Theoretical Physics, 1015 Lausanne, Switzerland}

\author{G.~Foffi}
\affiliation{Ecole Polytechnique F\'ed\'erale de Lausanne (EPFL), Institute of Theoretical Physics, 1015 Lausanne, Switzerland}

%\address[A. One and A. Two]
%{Author OneTwo common address, line 1 \newline%
%\indent Author OneTwo common address, line 2}%
%%%\email[F. Varrato]{francesco.varrato@epfl.ch}%
%%%\urladdr{http://personnes.epfl.ch/francesco.varrato}
%\author{Author Two}
%\curraddr[A.~Two]{Author Two current address, line 1\newline%
%\indent Author Two current address, line 2}%
%\email[A.~Two]{author-two@authortwo-inst.hu}%
%\urladdr{http://www.authortwo.uni-atwo.hu}
%\author{Author Three}
%\address[A. Three]{Author Three address, line 1\newline%
%\indent Author Three address, line 2}
%\email[A.~Three]{author-three@authorthree-inst.edu}%
%\urladdr{http://www.authorthree.uni-athree.edu}
%\thanks{Thanks for Author One.}
%\thanks{Thanks for Author Two.}

%\thanks{...}
%\date{January 11, 2010}
%\keywords{Physical Fractal, Fractal Dimension, Apollonian Packing, Algorithm}
%\dedicatory{Dedicated to The Great Geometer, in occasion of this \textit{n}th overuse of his name\ldots}
%
%\author{Francesco Varrato$^\ast$\thanks{$^\ast$Email: francesco.varrato@epfl.ch \vspace{6pt}} and Giuseppe Foffi
%\\\vspace{6pt} {\em{Ecole Polytechnique F\'ed\'erale de Lausanne (EPFL),\\Institute of Theoretical Physics, 1015 Lausanne, Switzerland}}\\\vspace{6pt} %\received{v2.0 released August 2011} 
%}

%\maketitle

\begin{abstract}
%	Corrections due to the finite size effect have been analytically introduced in the usual power-law distribution for fractal objects. This results in a generalization of the scale-free distribution and allows a quantitative study for finite ranges of sizes $s\in[s_\mathrm{min},s_\mathrm{max}]$, the \textit{physical fractals}, and not only for the asymptotic limit $s_\mathrm{min}/s_\mathrm{max}\rightarrow 0$. Moreover, a new efficient space-filling algorithm has been developed which generates osculatory random Apollonian packings (AP) of spheres with a finite range of diameters: not only the known AP's fractal dimensions are recovered but an excellent agreement with the generalized law is proved to be valid whitin the overall ranges of sizes.\\
	
	The Apollonian packings (APs) are fractals that result from a space-filling procedure with spheres. We discuss the finite size effects for finite intervals $s\in[s_\mathrm{min},s_\mathrm{max}]$ between the largest and the smallest sizes of the filling spheres. We derive a simple analytical generalization of the scale-free laws, which allows a quantitative study of such \textit{physical fractals}. To test our result, a new efficient space-filling algorithm has been developed which generates %osculatory
	random APs of spheres with a finite range of diameters: the correct asymptotic limit $s_\mathrm{min}/s_\mathrm{max}\rightarrow 0$ and the known APs' fractal dimensions are recovered and an excellent agreement with the generalized analytic laws is proved within the overall ranges of sizes.
	\\
\end{abstract}

%\pacs{XXX}
\maketitle

\begin{figure}[h!]
	\centering
		\includegraphics[width=8.5cm]{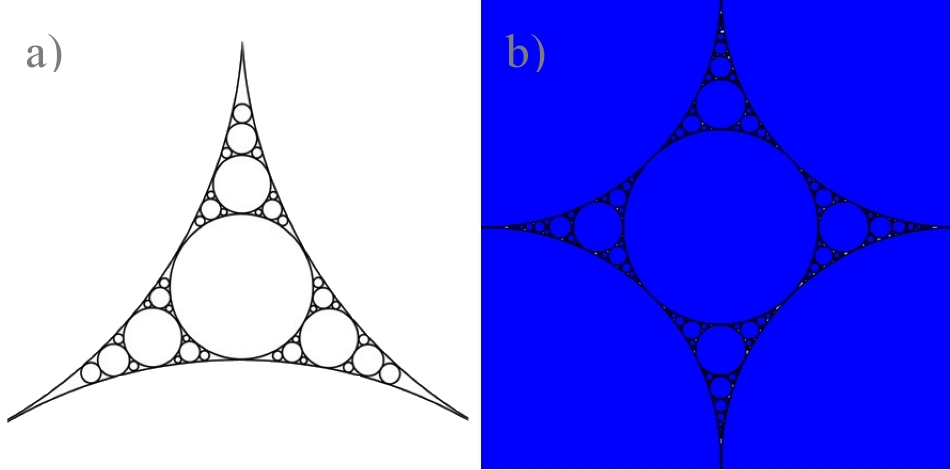}
	\caption{Apollonian packings for $d=2$: a) classical example as in Ref.~\cite{Kasner}; b) an example of random packing from the algorithm ($s_{\mathrm{max}}/L=1$) as explained later in the text.}
	\label{fig:circles}
\end{figure}

%%%%%%%%%%%%%%%%%%%%%%%%%%%%%%%%%%%%%%%%%%%%%%%%%%%%%%%%%%%%%%%%
\noindent {\bf 1. Introduction}\\
%%%%%%%%%%%%%%%%%%%%%%%%%%%%%%%%%%%%%%%%%%%%%%%%%%%%%%%%%%%%%%%%
\noindent The problem of finding the circle inscribed into the interstices between mutually tangential circles and tangent to them (a so-called \textit{osculatory} packing), historically attributed to Apollonius of Perga, % and has been fruitfully reprised since the 16th century.
was solved by Descartes and independently rediscovered various times \cite{Soddy}. Leibniz pointed out the possibility of obtaining a peculiar kind of packing by iterating the procedure of inserting such \textit{kissing} circles, whose size decreases as the inserting procedure goes on \cite{Hirano}. By starting from an initial configuration and by recursively filling the space with the osculatory packing down to arbitrarily small diameters, the \textit{Apollonian packing} (AP) is constructed. In Fig.~\ref{fig:circles} two examples of AP are presented for Euclidean dimension $d=2$. The structure of an AP is \textit{self-similar} because it is repeated on different scales of observation. In general, the self-similarity can be exact or statistical and leads to a fractal. The main quantity characterizing a fractal structure is the \textit{fractal dimension}, $d_f$, which is a (Lipschitz) invariant of the set descending from the Hausdorff-Besicovitch (HB) measure definition and possibly differs from the Euclidean dimension $d$ \cite{Mandelbrot,Ruelle}.\\
%The fractal dimension is a measure on a set of geometrical objects, e.g. set of polydisperse circles, thinkable as their degree of irregularity and related to the idea of covering of the topological space: it can be seen as the capacity for the given set of filling the topological space. 
%The HB definition of fractal dimension makes use of ``balls'' as geometrical covering objects; in the following we will always speak about spheres instead, but the discussion is indeed valid for every topological dimension $d$.\\
The osculatory packing construction has been extended to $d>2$ and the study of systems of polydisperse spheres %(non-overlapping balls frontiers) 
is made in the attempt of understanding how their fractality affects some macroscopic observables and, by reverse, how their formation mechanism influences their fractality. In particular, in $d=3$, the AP has been proposed as a model for dense granular systems \cite{Anishchik}.
It has been also used in describing stress yielding properties of materials, for example in concretes \cite{Aste}, as well as in the study of seismic gaps or geological faults \cite{Baram}. Moreover the scale-free properties of the AP are of particular interest in the context of complex networks \cite{Andrade,Massen}. Recently AP of non-spherical objects  have also been studied \cite{Dodds2,Delaney}.\\ While the application and the characterization of the AP has been widely studied, there is no exact theoretical prediction for the value of $d_f$ in 2 and 3 dimensions and various techniques have been developed to build AP and numerically evaluate their $d_f$. For example, it has been calculated for the plane tiled with circles obtained by the circular inversion method \cite{Manna2}. In $d=3$, a generalized inversion algorithm has been adopted \cite{Borkovec}.\\ %These deterministic techniques are applied to an initial set, called a \textit{Soddy} set, of $d+2$ mutually tangent spheres (or circles). 
So far we have discussed the deterministic AP but it has been proved that its fractal nature emerges also when a random sequence of space-filling insertions is pursued. In this generalized model \cite{Manna2, Manna}, called random Apollonian packing (RAP), the circles are inserted one at time with center positions randomly chosen; after the insertion, the diameter is simply inflated until it touches a previously inserted circle. Extended models make use of simultaneously inflating circles. All these models, deterministic or not, are shown to have universal features belonging to a broader class of models called ``packing-limited growth'' \cite{Dodds}. The RAP relies on the fact that all the osculatory packings in a certain dimensionality must have the same fractal dimension \cite{Boyd}. Several routes to RAP has been devised; we suggest Refs.~\cite{Amirjanov,Dodds,Delaney} as a short review.

\noindent Both deterministic and random procedures, when studied numerically, are affected by finite-size effects. In the context of AP (as in its random counterpart), starting from a given configuration of equal spheres of diameter $s_\mathrm{max}$, the impossibility of having packings with arbitrarily small sizes means that the asymptotic limit
\begin{equation}\label{asympt}
	s_\mathrm{min}/s_\mathrm{max}\rightarrow 0
\end{equation}
(where $d_f$ is defined) will never be reached.\\
It would be interesting to systematically characterize the deviation from such a limit.
In this paper, we propose a simple solution to this problem and we find that, taking finite size effects into account, the fractal dimension remains well defined also for finite ranges $s\in[s_\mathrm{min},s_\mathrm{max}]$.
The paper is organized as follows. In the next two paragraphs,  we introduce the basic ideas and observables and we provide an analytic solution for the problem of evaluating $d_f$ for a finite size range.  In the following sections, we will introduce a new algorithm to construct \textit{osculatory random AP} in arbitrary dimension with the possibility of tuning the size ratio defined in eq.~\eqref{asympt}.
After having verified its correct asymptotic behaviour, we shall finally use this novel algorithm to test the proposed finite size correction.\\

%%%%%%%%%%%%%%%%%%%%%%%%%%%%%%%%%%%%%%%%%%%%%%%%%%%%%%%%%%%%%%%%
\noindent {\bf 2. Apollonian packings and physical fractals}\\
%%%%%%%%%%%%%%%%%%%%%%%%%%%%%%%%%%%%%%%%%%%%%%%%%%%%%%%%%%%%%%%%
\noindent In the rest of the paper, we refer to the hard spheres as the set of non-overlapping geometrical objects (also for dimensions different from $d=3$) and the size $s$ then refers to the diameter.
Given a packing of equal hard spheres which occupy a volume fraction $\phi$, the porosity $\upvarepsilon\equiv 1-\phi$ %(in $d=2$ and in $d=3$ there is a maximum that corresponds to the hexagonal and to the fcc arrangement, respectively \cite{Hales}),
can be decreased by filling the free interstices with smaller spheres; %: this packing falls under the definition of \textit{packing-limited growth};
by iterating the insertion procedure of smaller spheres, the final packing results to be a fractal as $ s_{\mathrm{min}}$ goes to $0$, which corresponds to the limiting value $\phi=1$.
At equal number $N$ of total inserted objects, the Apollonian packing (AP) is known to be the densest of these packings \cite{Aste}. Despite the universality of $d_f$ for all osculatory packings \cite{Boyd}, i.e. for all configurations in which any interstice is filled with the largest possible hard sphere, there is not yet an analytical expression of the AP's fractal dimension. The fractal dimensions $d_f=1.305684$ and $d_f=2.473946$ have been numerically calculated for the AP in $d=2$ and $d=3$ respectively \cite{Manna2,Borkovec}.\\
In general the relation $d-1\leq d_f\leq d$ is valid, as confirmed by numerical simulations \cite{Kinzel} and the AP scale-free nature is expressed by the size distribution
\begin{equation}\label{n}
	n(s)\propto s^{-(d_f + 1)}\,\,,
\end{equation}
a power-law defined for all the positive diameters $s$ \cite{Manna}.\\
Fractals are implicitly understood to be scale-free structures and eq.~\eqref{n} is a consequence of that. However,  when a finite range of sizes exists, the fractal is called a \textit{physical fractal} \cite{Martinez3,Lakhtakia}. In this case, the distribution of $s$ is limited to an interval $[s_{\mathrm{min}}, s_{\mathrm{max}}]$, where $s_{\mathrm{min}}$ and $s_{\mathrm{max}}$ are the smallest and the largest diameters of the packing, respectively. Various practical methods have been introduced to calculate the fractal dimension: the \textit{box-counting}, for example, is applied in the case of physical fractals \cite{Martinez, Williams} and to direct measurements of physical systems \cite{Valle}.\\
We use two typical observables, the inverse cumulative distribution (number of spheres with size greater or equal to $ s_{\mathrm{min}}$) and the porosity, which under the asymptotic condition \eqref{asympt} are respectively given \cite{Manna2} by
\begin{subequations}
\begin{align}
N( s_{\mathrm{min}})&=\sum_{ s_i> s_{\mathrm{min}}} 1 \propto  s_{\mathrm{min}}^{-d_f}\,\,,\label{N}\\
\upvarepsilon( s_{\mathrm{min}}) &= 1 - \sum_{ s_i> s_{\mathrm{min}}} s_i^d \propto  s_{\mathrm{min}}^{d-d_f}\,\,.\label{p}
\end{align}
\end{subequations}
These measures can be used to estimate the $d_f$ for packings of polydisperse hard spheres: for different occurrences of $s_{\mathrm{min}}$, the fractal dimension is computable as smaller $s_{\mathrm{min}}$ are considered in measuring the values \eqref{N} and \eqref{p}\cite{Aste}.
All the different methods for evaluating the $d_f$ of packings rely on the fact that the fractal dimension definition works only approaching the asymptotic condition \eqref{asympt}. This means that, for finite ranges of diameters, relevant deviations exist starting from $s_{\mathrm{min}} / s_{\mathrm{max}} \gtrsim 1/5$, as explicitly highlighted in Ref.~\cite{Anishchik}.\\

\par
\noindent {\bf 2.1 Finite size deviation}\\
The main idea is that the geometrical building rule itself, with its iterativity, defines the fractal-like behavior, while the finite interval of sizes influences only the quantity of objects used for the building. The self-similarity of a set then implies the existence of a set of similarities in the generation of the fractal, which is beyond the mere agreement of the value of the fractal dimension \cite{Martinez2}.\\
This idea can be quantitatively rendered. For a random AP where the osculatory packing constrain is respected at each insertion (building iterative rule) we expect that the value for $d_f$ will remain the same. Due to the fractal nature of AP and RAP the power-law \eqref{n} should hold and 
%\textbf{We argue that what really matters in defining the fractal-like behaviour is the iterative rule itself, %(no matter for the technical way the filling rule is satisfied) 
%while the finite sizes interval can possibly influence only the number of fractal objects of maximum size. 
%This comes from the observation that the self-similarity of a set implies the existence of a set of similarities that generates the fractal, which is beyond the mere agreement of the value of the fractal dimension \cite{Martinez2}. This idea can be quantitatively translated for a random AP where the osculatory packing constrain is respected at each insertion: starting from a configuration of spheres with equal diameter $ s_{\mathrm{max}}$, the space between spheres is filled with smaller spheres down to a size $ s_{\mathrm{min}}$, each time maximizing the occupied space.(???)}\\ % by using only the simple fractal-like distribution instead of the HB measure definition, eq. 
we make the ansatz that the finite size correction is completely accounted for the proportionality constant of the distribution. In the finite case, we rewrite $n(s)$ as $n_f(s)$:
\begin{equation}\label{n2}
	n_f(s) \equiv f( s_{\mathrm{min}}, s_{\mathrm{max}}) s^{-(d_f + 1)}\,\,.
\end{equation}
The recursive fractal construction is accounted by the power-law and we now calculate the corrections to eqs.~\eqref{N} and \eqref{p} through the use of $f(s_{\mathrm{min}}, s_{\mathrm{max}})$.\\
The fraction of space occupied by $N_{s_{\mathrm{max}}}$ spheres of maximum size is by definition
\begin{equation}
	\phi_{ s_{\mathrm{max}}} = 1 - \upvarepsilon_{s_{\mathrm{max}}} = N_{ s_{\mathrm{max}}} \frac{C_d  s_{\mathrm{max}}^d}{V}\,\,,
\end{equation}
where $V$ is the total volume and the curvature factor is
\[
	C_d = \frac{2 \pi^{d/2}}{\Gamma(d/2)d} = \left\{ \begin{array}{l}
	\frac{1}{2^d}\frac{\pi^{d/2}}{(d/2)!}\,\,\,\textrm{\hspace{36pt}for even}\,d\\
	\frac{1}{2^{(d-1)/2}}\frac{\pi^{(d-1)/2}}{d!!}\,\,\,\textrm{\hspace{5pt}for odd}\,d
	\end{array}\right.\,\,.
\]
The total number of spheres $N$ and volume fraction $\phi$  can be expressed using the $0-$th and $d-$th moments of the distribution \eqref{n2} as:
\begin{subequations}
\begin{align}
	N &\equiv N_{ s_{\mathrm{max}}} + \int_{ s_{\mathrm{min}}}^{ s_{\mathrm{max}}} n_f( s') \mathrm{d} s'\,\,,\label{N2}\\
	\phi &\equiv \phi_{ s_{\mathrm{max}}} + \left(C_d/V\right)\int_{s_{\mathrm{min}}}^{s_{\mathrm{max}}} n_f( s')  s^d \mathrm{d} s'\,\,.\label{p2}
\end{align}
\end{subequations}
A natural condition arises assuming that all the space shall be occupied as the filling procedure continues; this means that, for any positive value of $s_{\mathrm{max}}$, it is
\begin{equation}\label{1}
\lim_{ s_{\mathrm{min}}\rightarrow0}\phi = 1\,\,.
\end{equation}
Using eq.~\eqref{n2} into eq.~\eqref{p2} with the limit \eqref{1}, we obtain the distribution's proportionality constant as
\begin{equation}\label{f}
	f( s_{\mathrm{min}}, s_{\mathrm{max}}) = N_{ s_{\mathrm{max}}} \frac{\upvarepsilon_{ s_{\mathrm{max}}}}{\phi_{ s_{\mathrm{max}}}}(d - d_f) s_{\mathrm{max}}^{d_f}\,\,.
\end{equation}
It is important to notice that it does not depend on $s_{\mathrm{min}}$. Inserting this result into eqs.~\eqref{N2} and \eqref{p2} finally gives:
\begin{subequations}
\begin{align}
	\frac{N}{N_{ s_{\mathrm{max}}}} &= 1 + \frac{(d-d_f)}{d_f}\frac{\upvarepsilon_{ s_{\mathrm{max}}}}{\phi_{ s_{\mathrm{max}}}} \left[\left(\frac{s_\mathrm{min}}{s_\mathrm{max}}\right)^{-d_f} - 1\right] \,\,,\label{fN}\\
	\upvarepsilon &= \upvarepsilon_{ s_{\mathrm{max}}} \left(\frac{s_\mathrm{min}}{s_\mathrm{max}}\right)^{d-d_f}\,\,.\label{fp}
\end{align}
\end{subequations}
Note that eqs.~\eqref{N} and \eqref{p} are the particular asymptotic cases of their more general expressions \eqref{fN} and \eqref{fp}, as expected.
These results remain valid in the limit $V\rightarrow\infty$.
Despite the simple hypothesis made, now we have an explicit expression for the  observables which allows to evaluate the deviation from the ideal case \eqref{asympt}.\\

\begin{figure}[h]
	\centering
		\includegraphics[width=7.5cm]{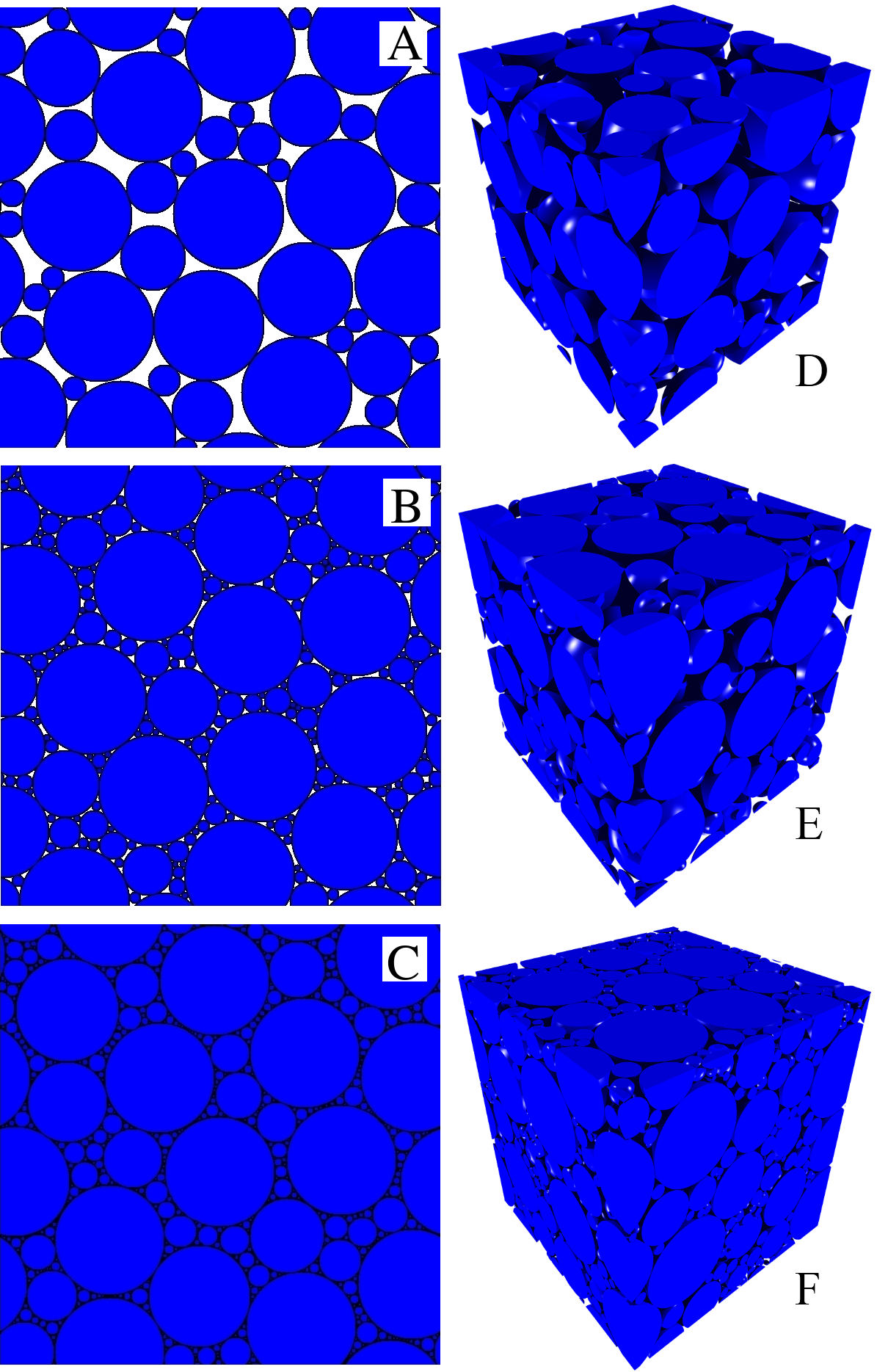}
	\caption{Snapshots of packings from independent algorithm runs: for $d=2$ $\textbf{(left)}$ with $s_\mathrm{min}/s_\mathrm{max}$ = 1/5, 1/20, 1/50 for A, B and C respectively (and the same $s_\mathrm{max}/L=1/4$); for $d=3$  $\textbf{(right)}$ with $s_{\mathrm{min}}/s_{\mathrm{max}}$ = 4/10, 2/10, 1/10 for D, E and F respectively (and the same $s_{\mathrm{max}}/L=1/3$).}
	\label{fig:Algorithm}
\end{figure}

%%%%%%%%%%%%%%%%%%%%%%%%%%%%%%%%%%%%%%%%%%%%%%%%%%%%%%%%%%%%%%%%
\noindent {\bf 2.2 Filling algorithm}\\
%%%%%%%%%%%%%%%%%%%%%%%%%%%%%%%%%%%%%%%%%%%%%%%%%%%%%%%%%%%%%%%%
In order to test the previous results, AP have been generated with the help of a new numerical algorithm (which works in any Euclidean dimension $d\geq1$). The developed algorithm has the same basic behavior as the \textit{random Apollonian packing} (RAP) mechanism, where the filling process starts with an initial population of hard-spheres of a specified diameter ($ s_{\mathrm{max}}$) and proceeds with new spheres added one at a time into the unoccupied space; randomly fixing the center of any new sphere, its size is determined by extending the diameter until it touches its closest sphere. But the RAP is not expected to build osculatory packings at each insertion, as we demand in order to check the theoretical previsions. In the algorithm we propose, random movements are additionally performed by the inserted spheres, in order to enhance the filled space accepting only the displacements that allow its diameter to grow (possibly up to $s_{\mathrm{max}}$).\\
No distribution of sizes for the spheres to insert is \textit{a priori} chosen, nor any initial population. Only the length-scales $s_{\mathrm{min}}/ s_{\mathrm{max}}$ and $s_{\mathrm{max}}/L$ are the parameters to be decided, where $L$ defines the total volume $V\equiv L^d$ of an initially empty box. Periodic boundary conditions are used so that possible interactions with walls can be ignored. In deterministic algorithms, the $d+1$ ``appropriate'' first neighbours must be identified so that the center's coordinates of a new sphere could be calculated (Soddy's rule); instead this random approach only has the non-overlapping constraint, meaning that the overall computational complexity is decreased.\\
A regular squared mesh with lattice constant $a_\mathrm{lattice}$ is defined into $V$ and its nodes are used as starting centers for the spheres to be inserted. The only requirement on the lattice constant is to be sufficiently smaller than $s_{\mathrm{min}}$. A fixed lattice constant could be preferred in some cases (the computational cost depends of course on the implementation); we tested $a_\mathrm{lattice} = s_{\mathrm{min}}/3$ to be a good parameter, but a recursive remeshing has been preferred and used to assure the requirement to be fulfilled.\\
As the first sphere is inserted at random on a starting center, it doesn't encounter any other spheres and its diameter can be expanded to $s_{\mathrm{max}}$. The starting centers it will cover will then be erased. One sphere at a time is subsequently inserted, according to the following scheme:
\begin{enumerate}
	\item[1:] it is placed at random on one of the remaining starting centers (nodes);
	\item[2:] its diameter is increased up to $s_{\mathrm{max}}$ or until it touches a previously inserted sphere;
	\item[3:] a random displacement within a maximum length $\Delta r\ll s_{\mathrm{min}}$ (average displacement $\Delta r/2$) is accepted only if this lets the diameter to grow;
	\item[4:] step 3 is iterated if the newly calculated diameter is $s < s_{\mathrm{max}}$;
	\item[5:] if $s_{\mathrm{max}}$ is reached or if the maximum number of displacement attempts $(s_{\mathrm{max}}/\Delta r)^2$ is reached, then the procedure stops.
	\item[6:] once a sphere has been inserted, the starting centers it covers are erased.
\end{enumerate}
The stopping conditions 5 rely on the possible random walk a sphere would need to explore a space of size $s_{\mathrm{max}}$. A maximum displacement length $\Delta r = |\Delta \vec{r}| \ll s_{\mathrm{min}}$ gives an average attempted displacement $\Delta r/2$ and the number $n_a$ of attempted displacements has been chosen to be the constant
\[
	n_a = \left( \frac{s_{\mathrm{max}}}{\Delta r} \right)^2
\]
corresponding to the number of steps for a random walk to explore a region of diameter $s_{\mathrm{max}}$. Of course a higher $n_a$ could result in a higher final packing fraction, but our calculation show poor improvement coupled with higher computational cost. During the procedure, spheres with final $s<s_{\mathrm{min}}$ are erased. The overall filling procedure ends when no other starting centers are present. In Fig.~\ref{fig:Algorithm} snapshots are shown for different $s_{\mathrm{min}}/ s_{\mathrm{max}}$ in $d=2$ and $d=3$. For our simulations we used the value $L=1$, the others lengths being consequently defined with respect to it. Once a value is chosen for $s_{\mathrm{max}}/ L$, the ratio $s_{\mathrm{min}}/ s_{\mathrm{max}}$ is investigated.
\\

\begin{figure}[h!]
	\centering
		\includegraphics[width=7.5cm]{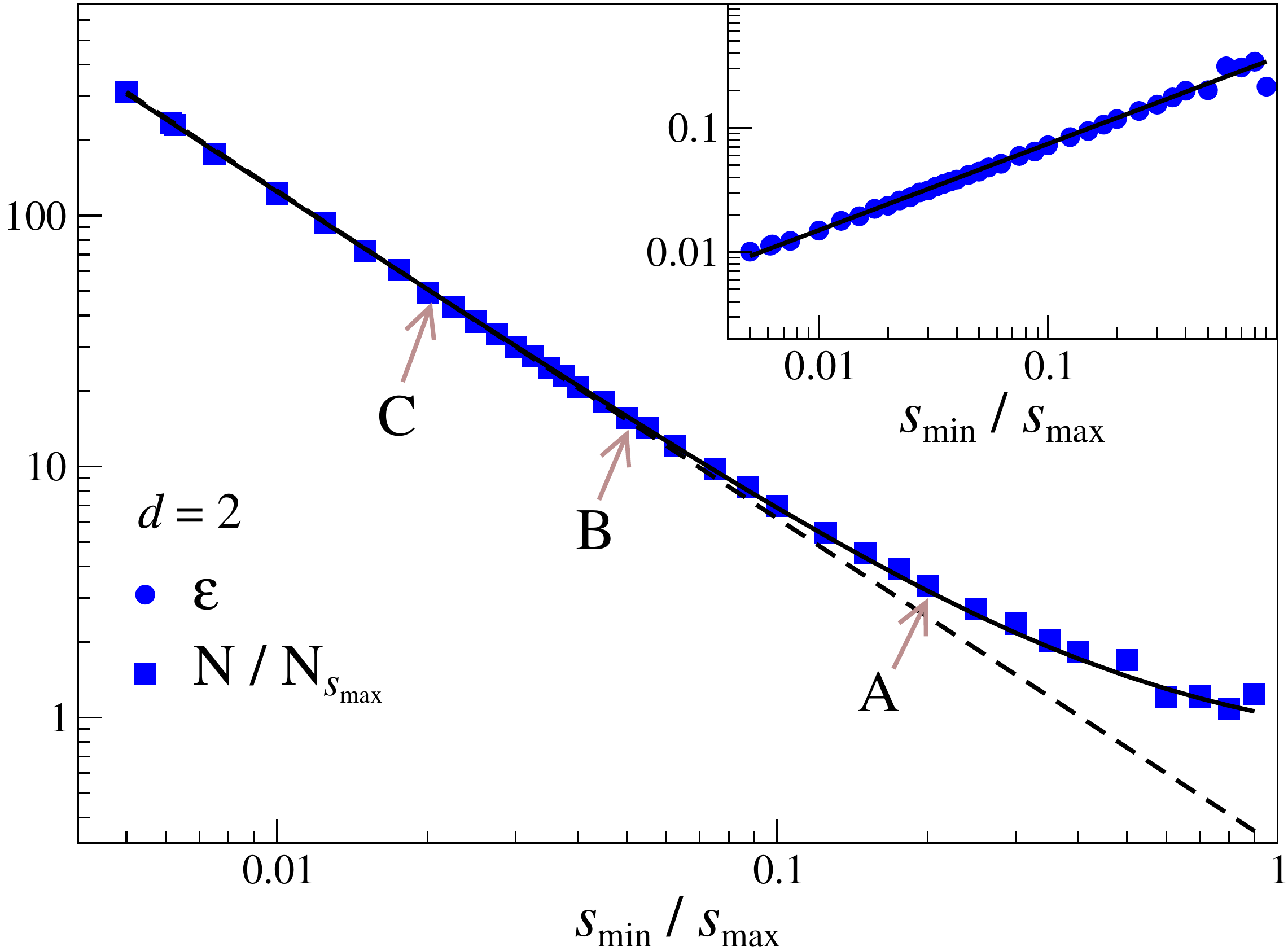}
		\includegraphics[width=7.5cm]{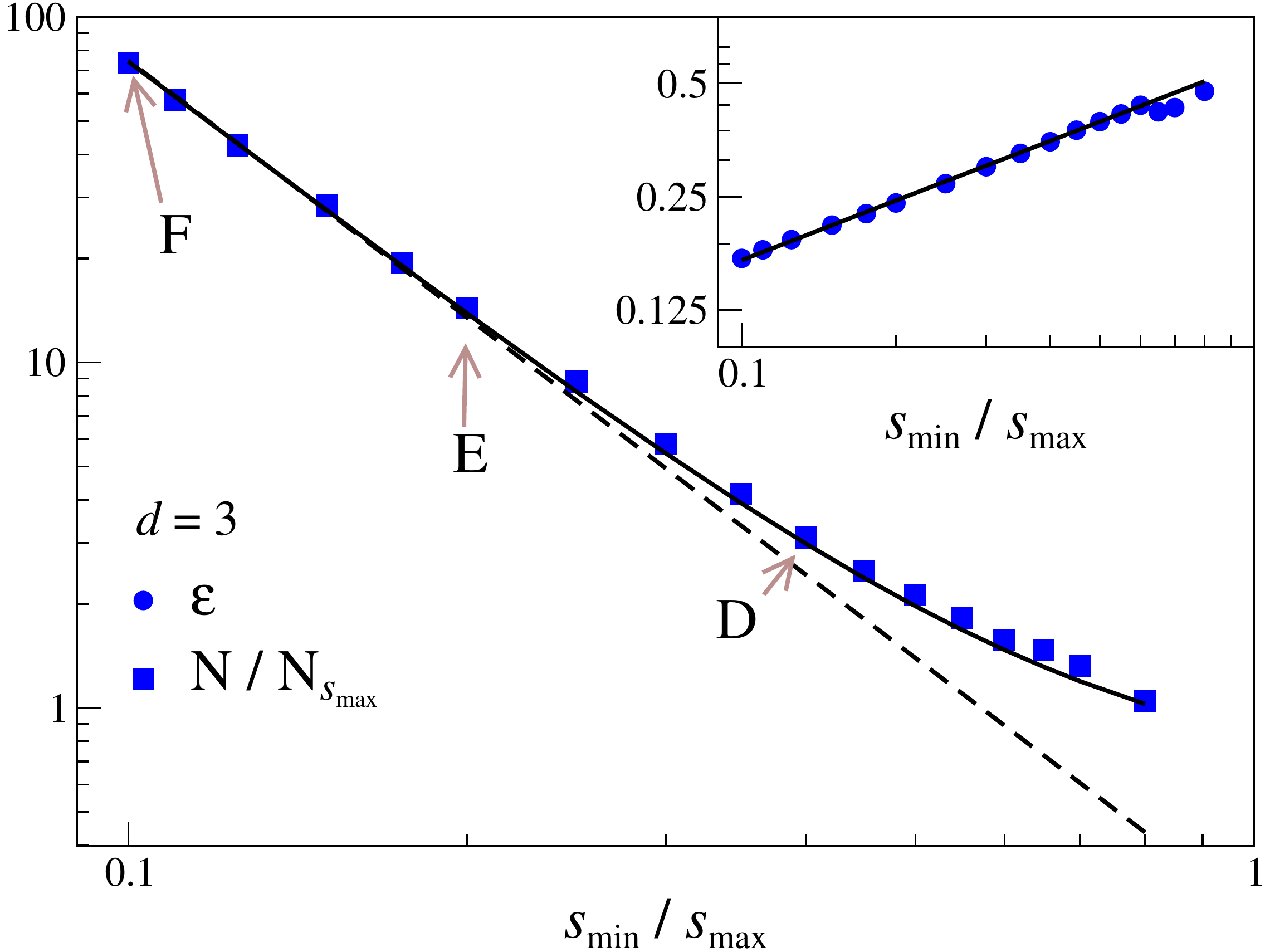}
	\caption{Number of inserted spheres $N$ (over $N_{s_{\mathrm{max}}}$) as function of the scale parameter $s_{\mathrm{min}}/s_{\mathrm{max}}$ for $d=2$ ($\textbf{top}$, $s_{\mathrm{max}}/L=1/5$) and $d=3$ ($\textbf{bottom}$, $s_{\mathrm{min}}/ L=1/3$). Each symbol represents the average over $10$ independent realisations of our random AP. The dashed lines represent eq.~\eqref{N}, while the solid lines correspond to the fits made using \eqref{fN}. Porosities $\upvarepsilon$ for the same systems are shown in the respective insets: the curves from eqs.~\eqref{p} and \eqref{fp} coincide with a power law. The positions for samples A, B, C, D, E and F in Fig.~\ref{fig:Algorithm} are indicated on the curves.}
	\label{fig:Results}
\end{figure}

%%%%%%%%%%%%%%%%%%%%%%%%%%%%%%%%%%%%%%%%%%%%%%%%%%%%%%%%%%%%%%%%
\noindent {\bf 2.3 Results}\\
%%%%%%%%%%%%%%%%%%%%%%%%%%%%%%%%%%%%%%%%%%%%%%%%%%%%%%%%%%%%%%%%
\noindent For the different values of $s_{\mathrm{max}}$ we have tested, we observed the same behaviour for decreasing $s_{\mathrm{min}}$; in the next paragraph we report on the results for 370 runs performed with $s_{\mathrm{max}}=1/5$ in $d=2$ and for 170 runs with $s_{\mathrm{max}}=1/3$ in $d=3$. These results are shown in Fig.~\ref{fig:Results}, where each symbol represents the value averaged over 10 independent runs.\\
For any configuration of $N_{s_{\mathrm{max}}}$ non-overlapping spheres in the total volume $V$, there always exist some values $d_f<d$ for which the condition \eqref{1} is satisfied. This does not imply the existence of a method capable of filling volume accordingly to \eqref{fN} and \eqref{fp}, but simply implies that if such an ``iterative method'' exists, then it allows the space to be occupied with a certain $d_f$. If this method consists in recursively filling the voids, each time maximizing the occupied space, then it should always present the same fractality as an AP. We expect that with such a kind of filling, including the algorithm presented here, not only  the correct asymptotic power-law behaviour has to be obtained, but also that the more stringent expressions \eqref{fN} and \eqref{fp} are fulfilled in the overall range of sizes. While power-laws \eqref{N} and \eqref{p} are known to work for $s_{\mathrm{min}}/ s_{\mathrm{max}}$ smaller than $1/5$, eqs.~\eqref{fN} and \eqref{fp} are in fact expected to work in the whole interval $s_{\mathrm{min}}/ s_{\mathrm{max}}\in[0,1]$.\\
We begin by testing the asymptotic behaviour of our osculatory RAP on the power law presented in eq.~\eqref{n} with the  values of $d_f$ for $d=2$ and $d=3$ obtained from previous calculations on AP \cite{Manna2,Borkovec}. As can be argued from Fig.~\ref{fig:Results}, in both cases good agreement exists for low enough values of $s_{\mathrm{min}}/ s_{\mathrm{max}}$, proving the correct AP asymptotic behaviour for the data obtained by the algorithm.\\
The numerical results deviate from the fractal asymptotic regime above certain size ratios, where we expect instead that our equations should still hold. To this aim we fit the data points for $d=2$ and $d=3$ with eqs.~\eqref{fN} and \eqref{fp} in the full range of sizes. It is important to stress that the value $N_{s_\mathrm{max}}$, which enters the definition of $N/N_{s_{\mathrm{max}}}$ and $\upvarepsilon_{ s_{\mathrm{max}}} = 1 - \phi_{ s_{\mathrm{max}}}$,  is not a fitting parameter. In fact it is known as the average number of spheres of diameter $s_\mathrm{max}$ in the obtained packings.\\
Results are shown in Fig.~\ref{fig:Results}; a comparison with the values known in the literature is reported in Tab.\ref{table}. The deviation from the asymptotic power-law is evident for $N/N_{s_{\mathrm{max}}}$, while the porosity holds its power law form ~\eqref{p} in eq.~\eqref{fp}.\\
\noindent The $N/N_{s_{\mathrm{max}}}$ curve results to describe the data better then its asymptotic counterpart. The fits finally allow an independent estimation of the fractal dimension $d_f$: our fitted fractal dimensions are in a very good agreement with the known values.  Note that the possibility of fitting in the whole $s$-range, plus the use of independent simulations, allows to evaluate $d_f$ by simulating a relatively small number of spheres.
%%%%%%%%%%%%%%%%%%%%%%%%%%%%%%%%%%%%%%%%%%%%%%%%%%%%%%
\setcounter{table}{0}
\begin{table}[h!]
%	\begin{center}
		\begin{tabular}{c c c}
		\hline
			Dimension ~~~& $d_f$ \cite{Manna2,Borkovec} ~~~& $d_f$ (fit) \\
		\hline
			$d=2$ ~~~& 1.305684 ~~~& 1.3045 $\pm$ 0.0006 \\ % rms%: 0.046
%		\hline
			$d=3$ ~~~& 2.473946 ~~~& 2.4739 $\pm$ 0.0014 \\ % rms%: 0.057
		\hline
		\end{tabular}
%	\end{center}
	\caption{values for the evaluated fractal dimension (in 2 and 3 Euclidean dimensions) compared with the known ones.}
	\label{table}
\end{table}\\
%%%%%%%%%%%%%%%%%%%%%%%%%%%%%%%%%%%%%%%%%%%
%\\

\noindent {\bf 3. Conclusions }\\
We have studied the properties of space filling packings of spheres. In particular we have derived the finite size correction to the distribution laws that characterize the Apollonian packing fractals. In the case of sizes limited to a finite interval, the ratio between the smallest and the largest spheres does not go to zero and deviations are indeed expected from the typical power laws. 
Two main observables have been studied, the ratio of the total number of spheres over the number of largest spheres, i.e. $N/N_{s_{\mathrm{max}}}$, and the fraction of unoccupied space, i.e. the porosity $\upvarepsilon$. We provide simple analytical expressions for them, solely based on the hypothesis that the packed spheres totally occupy the space if the minimum diameter tends to zero.\\
In order to test our prediction, an efficient algorithm has been introduced to generate osculatory random AP, not based on  any \textit{a priori} size distribution. This algorithm allows to fix, as input parameters, the largest and smallest sizes. In the limit of vanishingly small diameters,  disordered Apollonian packings are recovered with the correct asymptotic behaviour, as proved by testing the data obtained from the new packing algorithm. 
The laws corrected for finite sizes have been tested  by varying the interval of sizes for fractal objects (circles in $d=2$ and spheres and $d=3$) and the result of the fits allows to verify the values of the fractal dimension which came out in agreement with the values known from the literature.\\
It is interesting to note that the laws we derived apply to the whole range of size ratios. This suggests that even in the case of packings with a very narrow interval of sizes, the space filling construction preserves its fractal nature. 
These simple results could be applied to the broad class of ``packing-limited growth'' models and physical fractals for which the general conditions \eqref{n2} and \eqref{1} are valid. We finally propose to use the rapid convergence to an osculatory packing enhanced with the proposed algorithm, together with the possibility of using the whole range of sizes for the evaluation of the fractal dimension, as a possible feasible test on recent studies on Apollonian gaskets at higher dimensionality \cite{Farr}.
%\newpage

\bibliographystyle{apsrev}

\begin{thebibliography}{9}                                                                     


\bibitem{Soddy} F. Soddy, \textit{Nature} \textbf{137}, 1021 (1936)

\bibitem{Hirano} H. Hirano, \textit{``Leibniz's Cultural Pluralism And Natural Law''}, Hosei University, Tokyo (2010)

\bibitem {Mandelbrot} B. B. Mandelbrot, \textit{Fractal Geometry of Nature} (1983) San Francisco: Freeman Ed.

\bibitem {Ruelle} J. Fr\"{o}hlich and D. Ruelle, \textit{Scaling and Self-Similarity in Physics} (1983) Birkhauser Ed.

\bibitem{Anishchik} S. V. Anishchik and N. N. Medvedev, \textit{Phys. Rev. Lett.} \textbf{75} (1995), 4314–4317

\bibitem {Aste} T. Aste and D. Weaire, \textit{The pursuit of Perfect Packing} (2000) Bristol: Institute of Physics Publishing

\bibitem{Baram} R. M. Baram and H. J. Herrmann, \textit{Phys. Rev. Lett.} \textbf{92} (2004), 044301

\bibitem{Andrade} J. S. Andrade, Jr., H. J. Herrmann, R. F. S. Andrade and L. R. da Silva, \textit{PRL} \textbf{94} (2005), 018702

\bibitem{Massen} J. P. K. Doye and C. P. Massen, \textit{Phys. Rev. E} \textbf{71} (2005), 016128

\bibitem{Dodds2} P. S. Dodds and S. Weitz, \textit{Phys. Rev. E} \textbf{67} (2003), 016117

\bibitem{Delaney} G. W. Delaney, S. Hutzler and T. Aste, \textit{Phys. Rev. Lett.} \textbf{101} (2008), 120602

\bibitem {Manna2} S. S. Manna and H. J. Herrmann, \textit{J. Phys. A} \textbf{24} (1991), L481-L490

\bibitem {Borkovec} M. Borkovec, W. de Paris and R. Peikert, \textit{Fractals} \textbf{12} (1994), 521

\bibitem {Manna} S. S. Manna, \textit{Physica A} \textbf{187} (1992), 373

\bibitem{Dodds} P. S. Dodds and J. S. Weitz, \textit{Phys. Rev. E} \textbf{65} (2002), 056108

\bibitem {Boyd} D. W. Boyd, \textit{Can. J. Math.} \textbf{25} (1973), 303-322

\bibitem {Amirjanov} A. Amirjanov and K. Sobolev, \textit{Modelling Simul. Mater. Sci. Eng.} \textbf{14} (2006), 789-798

\bibitem {Kasner} E. Kasner and F. Supnick, \textit{PNAS} \textbf{29} (1943), 378-384

\bibitem {Kinzel} W. Kinzel and G. Reents, \textit{Physics by Computer} (1998) New York: Springer Ed.

\bibitem {Martinez3} F. Mart\'{\i}nez-L\'{o}pez, M. A. Cabrerizo-V\'{\i}lchez and R. Hidalgo-\'{A}lvarez, \textit{Physica A: Statistical Mechanics and its Applications} \textbf{298} (3-4) (2001), 387-399

\bibitem{Lakhtakia} A. Lakhtakia, \textit{Speculations in Science and Technology}, \textbf{18} (1995), 153-156

\bibitem {Williams} G. P. Williams, \textsl{Chaos Theory Tamed}, chapter \textbf{21}, (1997) London: Taylor\&Francis Ed.

\bibitem {Martinez} F. Mart\'{\i}nez-L\'{o}pez, M. A. Cabrerizo-V\'{\i}lchez and R. Hidalgo-\'{A}lvarez, \textit{Physica A: Statistical Mechanics and its Applications} \textbf{311} (3-4) (2002), 411-428

\bibitem{Valle} F. Valle, M. Favre, P. De Los Rios, A. Rosa and G. Dietler, \textit{Phys. Rev. Lett.} \textbf{95} (2995), 158105

\bibitem {Martinez2} F. Mart\'{\i}nez-L\'{o}pez, M. A. Cabrerizo-V\'{\i}lchez and R. Hidalgo-\'{A}lvarez, \textit{Journal of Physics A: Mathematical and General} \textbf{34}, 36 (2001), 7393-7398

\bibitem {Farr} R. S. Farr and E. Griffiths , \textit{Phys. Rev. E} \textbf{81} (2010), 061403

%\bibitem {Hales} T. C. Hales, \textit{Discrete Comput. Geom.} \textbf{18} (1997), 135

%\bibitem {Serrano} M. \'Angeles Serrano, Dmitri Krioukov and Mari\'an Bogun\'a, \textit{PRL} \textbf{106} (2011), 048701 

%\bibitem {Andrade} Jos\'e S. Andrade, Jr., Hans J. Herrmann, Roberto F. S. Andrade and Luciano R. da Silva, \textit{PRL} \textbf{94} (2005), 018702

%\bibitem {AndradeCorr} Jos\'{e} S. Andrade, Jr., Hans J. Herrmann, Roberto F. S. Andrade and Luciano R. da Silva, \textit{PRL} \textbf{102} (2009), 079901
%
%\bibitem {Mungan} Muhittin Mungan, \textit{PRL} \textbf{106} (2011), 029802

%\bibitem {Liu} J. Liu and K. Regenauer-Lieb, \textit{Phys. Rev. E} \textbf{83} (2011), 016106

%\bibitem {Kaplan} C. Nadir Kaplan, Michael Hinczewski and A. Nihat Berker, \textit{Phys. Rev. E} \textbf{79} (2009), 061120 

%\bibitem {Newman} M. E. J. Newman and R. M. Ziff, \textit{Phys. Rev. E} \textbf{64} (2001), 016706 
%

\end{thebibliography}

\end{document}